\documentclass[12pt,preprint]{aastex}

\shorttitle{X-ray Observations of H1426}
\shortauthors{Falcone, Cui, & Finley}

\begin{document}


\title{X-Ray Spectral Variability of Extreme BL Lac AGN H1426+428}


\author{A.D. Falcone\altaffilmark{1}, W. Cui, and J.P. Finley}
\affil{Physics Department, Purdue University, West Lafayette, IN 47901}


\altaffiltext{1}{located at Whipple Observatory, P.O. Box 97, Amado, AZ 85645}


\begin{abstract}
Between 7 March 2002 and 15 June 2002, intensive X-ray observations were carried out on the extreme BL Lac object H1426+428 with instruments on board the Rossi X-ray Timing Explorer (RXTE).  These instruments provide measurements of H1426+428 in the crucial energy range that characterizes the first peak of its spectral energy distribution.  This peak, which is almost certainly due to synchrotron emission, has previously been inferred to be in excess of 100 keV.  By taking frequent observations over a four-month campaign, which included $\sim$450 ksec of RXTE time, studies of flux and spectral variability on multiple timescales were performed, along with studies of spectral hysteresis.  The 3-24 keV X-ray flux and spectra exhibited significant variability, implying variability in the location of the first peak of the spectral energy distribution.  Hysteresis patterns were observed, and their characteristics have been discussed within the context of emission models.
\end{abstract}


\section{Introduction}
Blazars, including flat spectrum radio quasars (FSRQs) and BL Lacs, represent the most extreme examples of active galactic nuclei (AGN) due to their rapid variability and the extreme energies to which their spectra extend.  One of the most exciting developments in extragalactic astrophysics during the last decade has been the detection of more than 65 AGN in the energy range from 30 MeV - 30 GeV \citep{har99} by the EGRET experiment on the Compton Gamma Ray Observatory (CGRO).  The detection of some AGN (e.g. Mrk 421, Mrk 501, 1ES 2344+514, PKS 2155-304 and most recently H1426+428 and 1ES1959+650) above 300 GeV by ground-based air cherenkov telescopes (ACTs) has reinvigorated the study of these objects \citep{cat99,hor02,hol03}.  The dominant radiation from blazars is widely believed to arise from relativistic jets that are viewed at small angles to their axes \citep{bla79}.  EGRET and Whipple observations of high fluxes above 1 GeV and 300 GeV (respectively), the short-term variability of the gamma-ray emission \citep{kni93,mat93,qui96,buc96}, and constraints on gamma-ray opacity due to pair-production with low energy photons are all pieces of evidence leading to the conclusion that the gamma-ray emission is also beamed and that jets are involved.\\

The broadband radiation spectrum of blazars consists of two parts: a synchrotron spectrum that spans radio to optical-ultraviolet wavelengths (and to X-rays for high-peaked objects) and a high-energy part that can extend from X-rays to gamma rays.  Observationally, the spectra appear to have two distinct bumps when plotted in a $\nu$$f_{\nu}$ representation \citep{fos98,ghi98}.  The emission in the first peak of the spectral energy distribution (SED) is generally considered to be synchrotron emission from relativistic electrons.  The most widely invoked models which attempt to explain the higher energy emission of the second SED peak fall into the category of leptonic models.  These leptonic models posit that the X-ray to gamma ray emission is produced by inverse Compton scattering of lower energy photons by relativistic electrons within a narrow jet.  The low energy photon fields could arise from synchrotron continuum photons within the jet (e.g. \citet{koe81}), or they could arise from ambient photons from the accretion disk which enter the jet directly (e.g. \citet{der92}) or after scattering or reprocessing (e.g. \citet{sik94}).  In addition to these leptonic models, hadronic models also attempt to explain the second bump of the SED.  These models involve proton-initiated cascades (e.g. \citet{man93}) and/or proton synchrotron radiation \citep{mue01, aha00}, as well as synchrotron emission from secondary muons and pions\citep{mue03}.\\

The gamma-ray and simultaneous X-ray observations of blazars have placed strong constraints on these models \citep{von95, mac95, buc96} by constraining the properties of the emission regions (size, Doppler boosting factor, magnetic field, electron cooling time, total injection energy, ...) and by comparing flaring timescales at different wavelengths; but so far the observations cannot definitively reject either hadronic or leptonic models of the high energy emission.  X-ray observations are particularly important for observations of high-peaked BL Lacs since the peak of the synchrotron component lies in the band covered by X-ray detectors (e.g. Beppo-SAX observations of Mrk 501 \citep{pia98} and H1426+428 \citep{cos01}). The determination of the X-ray peak is highly constraining to the suite of models for the emission from these objects, as are the timescales of the X-ray variability.\\

Further constraints to emission mechanisms may be obtained by exploring the characteristics of spectral hardness-intensity diagrams (HIDs) in the X-ray energy band, with the purpose of searching for hysteresis patterns.  The presence of hysteresis loops and the handedness of such loops can be used to constrain the cooling and acceleration/injection timescales of the relativistic electron population \citep{bot02, li00, kir98}.  Spectral hysteresis has been observed for several high-peaked BL Lac objects, including Mrk 421 and PKS 2155-304 \citep{tak96, kat00, zha02}.  Another set of observations of Mrk 421 using XMM reported no significant hysteresis \citep{sem02}, however this result appears to be inconclusive due to the lack of coverage throughout the entire cycle of flux rise and subsequent decay coupled with the low amplitude variation of the one complete rise and decay.  \\

H1426+428 is an X-ray selected high-peaked BL Lac with a relatively strong X-ray flux.  It was first characterized as a BL Lac, due to its lack of prominent optical emission lines, by \citet{rem89}.  Based upon observations by Beppo-SAX up to 100 keV, H1426+428 is an example of an extreme high energy peaked BL Lac (HBL) with the peak of the synchrotron emission lying above 100 keV \citep{cos01} during that observing campaign.  This fact led \citet{cos01} to speculate that H1426+428 is a prime candidate for TeV gamma-ray emission; a speculation recently confirmed by the Whipple collaboration \citep{hor02, pet02}. Characterizing the X-ray properties of this extreme object is important, especially due to the presence of this TeV emission in spite of a large redshift (z=0.129).  Previous X-ray observations presented by \citet{sam97} and \citet{mad92} using ASCA, ROSAT, and BBXRT (at different epochs separated by as much as 15 years) have shown evidence for spectral variability from one observation to the next, as well as evidence for spectral curvature and a spectral absorbtion feature at $\sim$0.6 keV.  The results presented here represent a fairly continuous and deep exposure obtained for the purpose of exploring the details of the X-ray variability on multiple timescales.

\section{Observations}
X-ray observations were carried out between 7 March and 15 June of 2002.  These observations were part of a multiwavelength campaign \citep{fal03}, including radio, optical, X-ray, and gamma-ray energy bands.  This report will describe the X-ray observations (multiwavelength observations will be reported on in a separate publication).  X-ray observations were performed by PCA and HEXTE on the RXTE satellite.  The results reported here are based on PCA data.  The HEXTE data, which can only be analyzed on significantly larger timescales due to sensitivity constraints, will be described in the forthcoming multiwavelength paper.  Typical RXTE exposures were between 1-4 hours per night for 3 weeks bracketing the new moon of each month during the 4 month long campaign.  This resulted in $\sim$450 ksec of RXTE observations.

\subsection{Instrument}

The PCA instrument is composed of five proportional counter units (PCUs) and has a total collection area of 6500 cm$^{2}$ when all counters are operating.  However, during the observations discussed in this paper, only 2-3 of the proportional counters were on during normal operation.  For nearly all of this campaign, the proportional counters known as PCU0 and PCU2 were collecting data, while the other PCUs were dormant.  For consistency during data reduction, a cut was placed on the data to select only the PCU0 and PCU2 data.  Data from these PCUs were extracted independently of one another and they were combined by jointly modelling the spectra at the time of analysis.  The best statistics were obtained by using the 1st Xenon layer of each PCU and excluding layers two and three.  Cuts were also made to eliminate data that included effects due to passage through the South Atlantic Anomaly, as well as data for which the electron rate was anomalously high.  An elevation cut ($>$10$\deg$) was also imposed.\\

The extraction and processing of the data was done using ftools 5.2.  A response matrix was calculated, using the latest available software (PCARSP V8.0), individually for each PCU with the same configuration used for the data extraction process.  The weak source background model provided by RXTE/GOF was used for background subtraction.  This model appeared stable and reproduced the observed spectrum at high energies, where the spectrum is expected to be background dominated.  It is worth noting that the background model worked well for PCU0, which no longer has an operating propane veto layer.  The model background for PCU0 converged to the observations at high energies and the rate variations in PCU0 mimicked those of PCU2 which has an operating propane veto layer.

\subsection{Analysis}
After extraction of the data, XSPEC 11.2.0 was used to analyze and apply models to the extracted spectra.  The nominal energy range of the PCA is 2 to $\sim$60 keV, but since the response of the instrument is poorly characterized in the lowest energy bins, only data above 2.9 keV is used.  Due to the combination of falling source spectra for H1426+42.8 and sensitivity as energy increases, there is no significant measurement above 24 keV.  Based upon these two bounds, all analysis is confined to the region between 2.9 and 24 keV, and all other energy bins are excluded.  The 3.8-5.4 keV energy region is also excluded from the analysis since there is a known Xenon absorbtion feature within this energy region that is not modelled well by the instrument response.  The spectra are then fit to data that are binned into 1$-$day bins, as described below, and the fits are evaluated using a $\chi^{2}$ analysis and an inspection of the residuals.  Count rates in particular energy regions are obtained prior to application of spectral models in order to evaluate variability in the lightcurve.  Flux is derived based upon the parameters of the model that fits the spectral data.

\section{Results}

\subsection{Flux and Flux Variability}

A lightcurve showing the X-ray rate as a function of time in different X-ray energy bands is shown in figure 1.  This data has been binned into one day time intervals beginning with the first day of the campaign, and all error bars are 1$\sigma$ statistical errors.  The 2.9-10 keV and the 10-24 keV energy bands have been evaluated independently of one another and plotted, along with the entire 2.9-24 keV band.  The 2.9-24 keV flux observed by PCA varied between about 1.3x10$^{-11}$ to 6.6x10$^{-11}$ erg cm$^{-2}$ s$^{-1}$, and the mean 2.9-24 keV flux was 3.6x10$^{-11}$ erg cm$^{-2}$ s$^{-1}$.
\\
Variability is evident in all of the three bands.  Based on this analysis, which uses bins of 1-day duration, variability studies are limited to the timescale range bracketed by 1-day and 4-months.  There appears to be long term variability as the flux changes throughout the entire campaign with timescales in excess of two weeks, and there also appears to be variability on timescales as short as 1-2 days.  While the variability was significant, especially on long timescales, there were no particularly large flares relative to those observed for other TeV emitting BL Lacs, such as Mrk 421, Mrk 501, and 1ES 1959+650.

\clearpage
\begin{figure}
\plotone{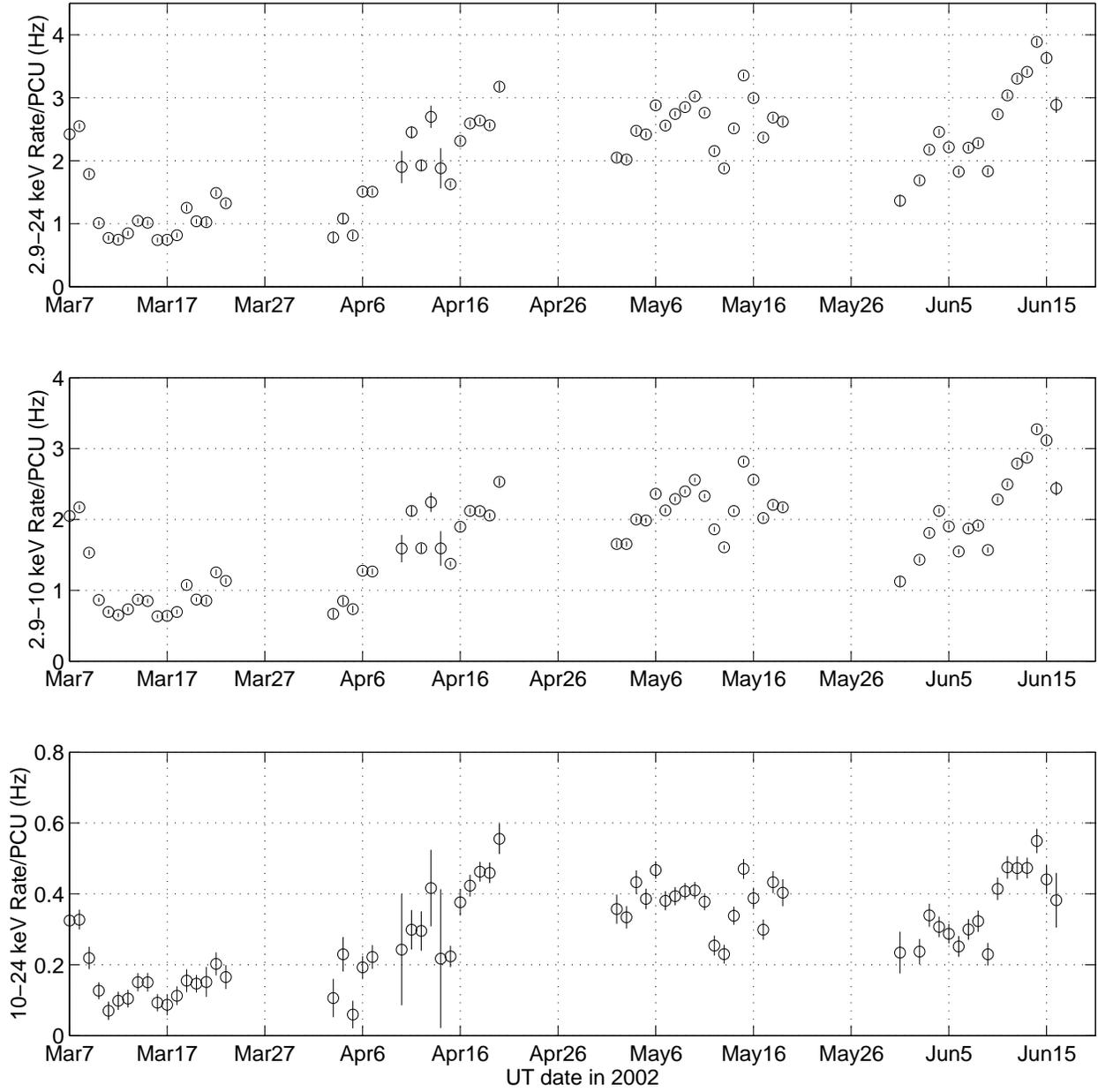}
\caption{X-ray count rate per PCU as a function of time during Mar-Jun of 2002 in three energy bands.  \label{fig1}}
\end{figure}
\clearpage
\subsection{Spectra and Spectral Variability}

For each set of data in a one day time interval, a model of the spectra was derived.  It was found that a power law with galactic absorbtion, column density $N_H = 1.4\times10^{20}$ \citep{sta92}, led to an excellent fit based upon a $\chi^{2}/dof$ analysis.  (In fact, rather than hovering at $\sim$1, the $\chi^{2}/dof$ is actually consistently less than one, due to the fact that the background rate error bars, which are added in quadrature to the source rate error bars, are conservatively large.  This occurs since the error bars on the background are assigned based on the counts in the observation, rather than the counts used to construct the background model, which is a sum of many observations with far more numerous counts.)  The power law spectra are of the form:
\begin{equation}
f = [C(\frac{E}{1 keV})^{-\Gamma}][e^{-N_{H}\sigma(E)}]
\end{equation}
where the units of the normalization constant C are photons $keV^{-1} cm^{-2} s^{-1}$, $\sigma(E)$ is the photo-electric absorbtion cross section given by \citet{mor83}, and $\Gamma$ is the power law spectral index.
Allowing the absorbtion column density to vary as a free parameter did not significantly effect the fit so it was fixed to the galactic value quoted above.  An example of a fit using this data is shown in figure 2. \\
\clearpage
\begin{figure}
\epsscale{0.60}{
\center{
\rotatebox{270}{
\plotone{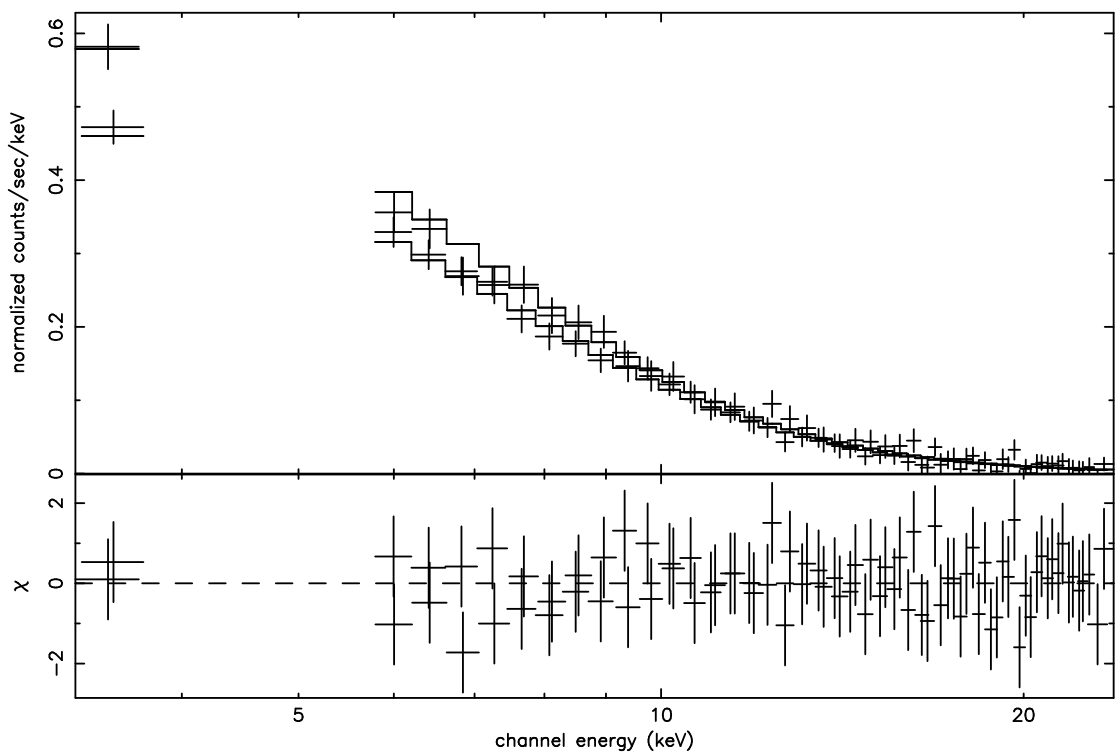}}
\caption{ An example of a model fit to a typical day in the middle of the campaign (6 May 2002), using the power law with galactic absorbtion model to jointly fit PCU0 and PCU2 data.  Residuals of the fit (in terms of sigma) are shown in the bottom panel.  \label{fig2}}}}
\end{figure}
\clearpage
The power law spectral index derived using the above procedure is shown as a function of time in figure 3, along with $\chi^{2}/dof$ for the particular fit.  All error bars are 1$\sigma$.  The error bars for the 2.9-24 keV flux were calculated by applying simple error propagation to the 1$\sigma$ error bounds produced by XSpec for the power law spectral index and normalization constant.  It can be seen from this plot that there was significant spectral variability during the campaign.  The spectral index varies over the range from 1.46$\pm$0.05 to 2.03$\pm$0.09.  The X-ray flux in the 2.9-24 keV band has also been derived based upon the spectra that were fit to the data.  This flux is shown in the top panel of figure 3.
\clearpage
\begin{figure}
\epsscale{1.00}
\plotone{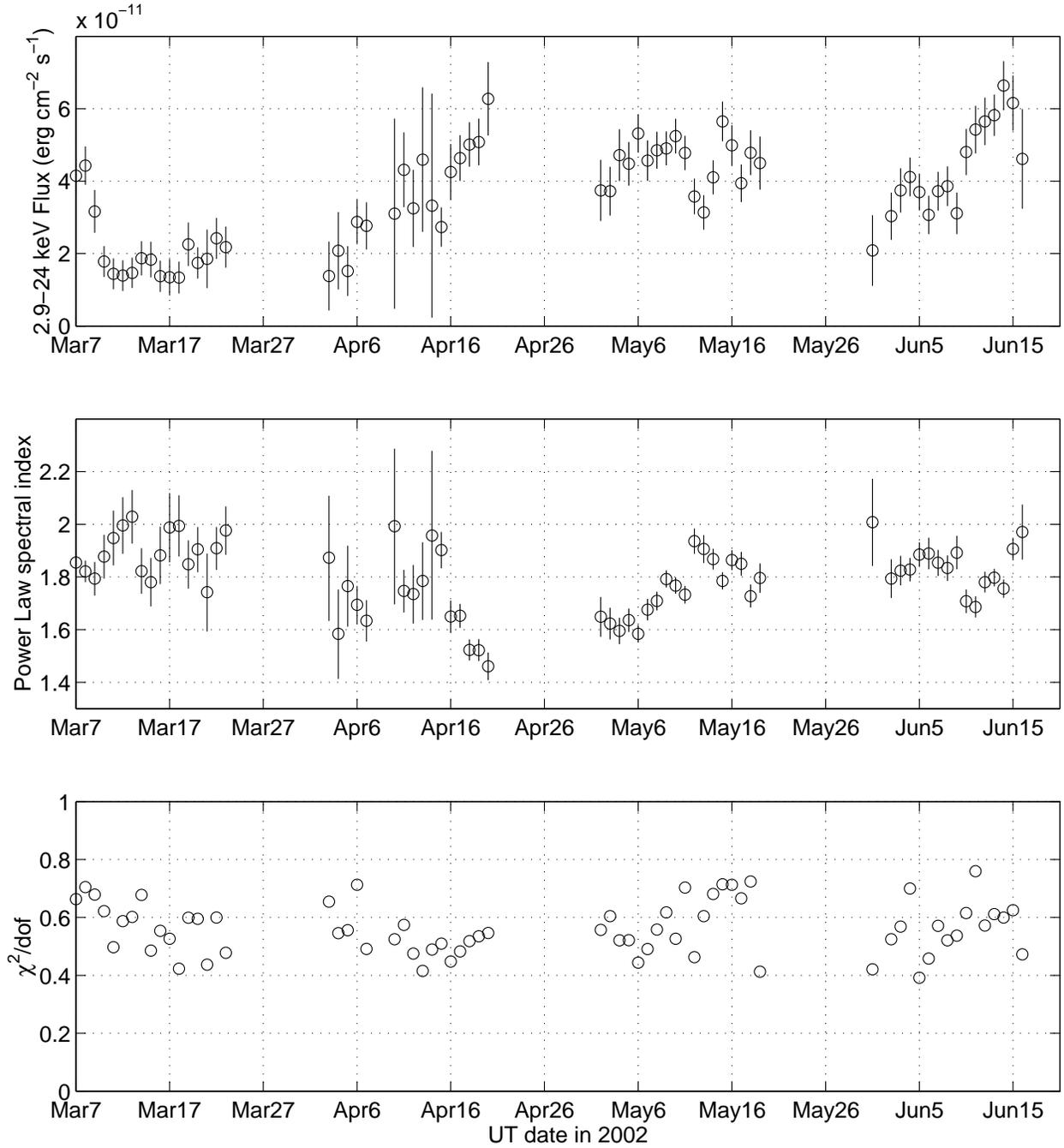}
\caption{Top panel: Xray flux as a function of time in the 2.9-24 keV band; bottom two panels: X-ray power law spectral index as a function of time along with the associated $\chi^{2}/dof$ for the fit.  Error bars are $1\sigma$. \label{fig3}}
\end{figure}
\clearpage
\subsection{Variability Correlations and Hysteresis}
Over the time frame of the entire campaign, there is no tight correlation between the X-ray flux and the power law spectral index.  The correlation coefficient between the 2.9-24 keV flux and the spectral index has been calculated using the 1-day timescale data throughout the entire four-month campaign, and its value is 0.51.  By simple inspection of the plots in figure 3, one can see that the spectral variability does not have a consistent relationship with the 2.9-24 keV flux variability over long timescales.  A plot of the spectral index vs. flux is shown in figure 4.  While, on average throughout the campaign, the flux increases as the spectrum becomes harder, there is a great deal of scatter on the HID for the entire campaign.  This plot includes flux variations averaged over long timescales so it is certainly not reflective of the source characteristics during a single flare event.  This lack of strong correlation does not preclude a relationship during individual flare timescales.
\clearpage
\begin{figure}
\plotone{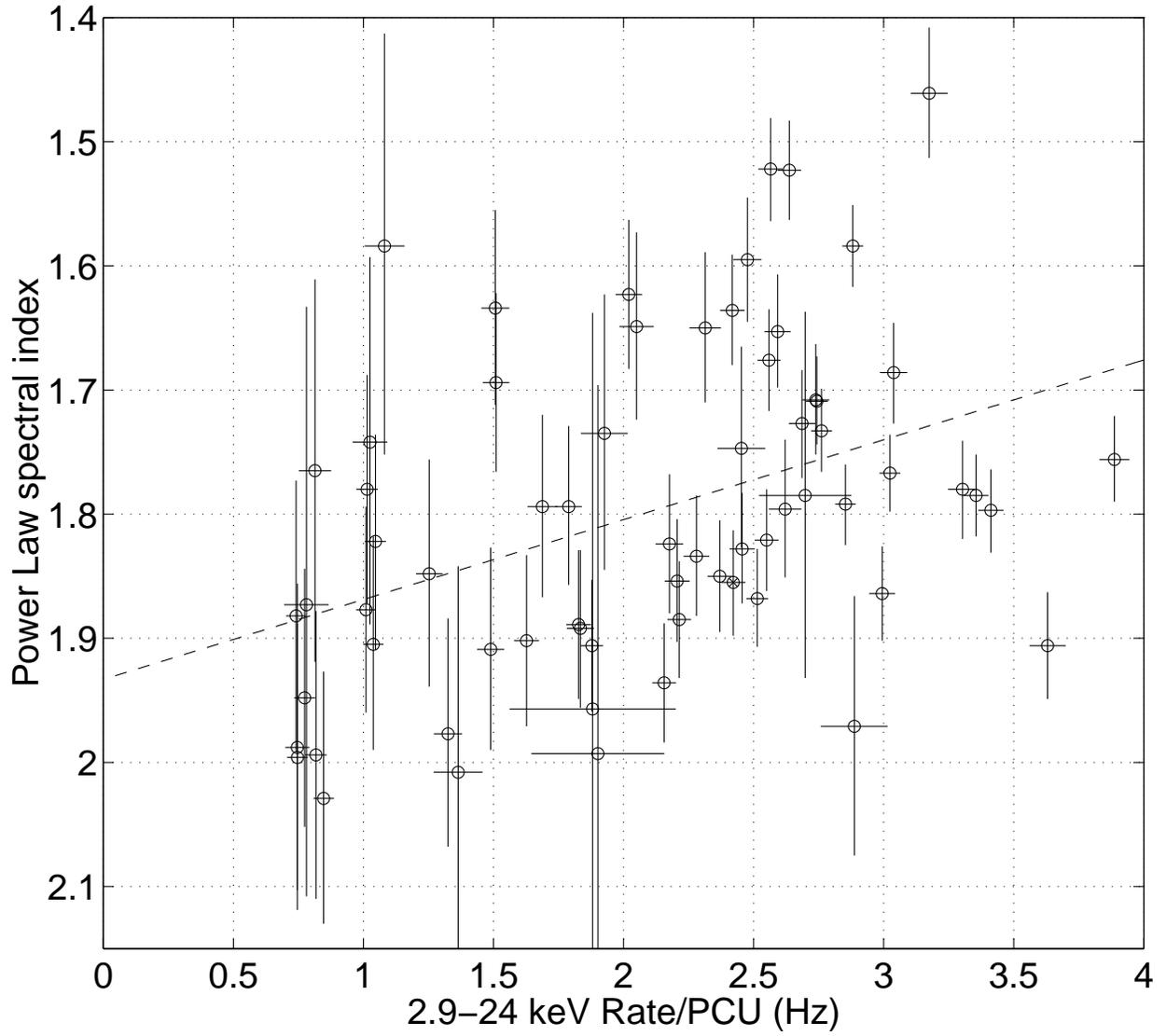}
\caption{Hardness-intensity diagram for the complete four-month long data set.  A linear fit to the data has been overlayed on the figure to guide the eye. \label{fig4}}
\end{figure}

\clearpage
\begin{table}
\begin{center}
\caption{Definitions of time regions during periods of flux variation \label{Table-1}}
\begin{tabular}{ccc}
\tableline\tableline
Region & Time & Characteristics \\
\tableline
A & 1-12 May & long rise, fast decay \\
B & 12-16 May & 2-day rise, 2-day decay \\
C & 30May-5Jun & 4-day rise, 2-day partial decay \\
D & 5Jun-15Jun & long rise, 2-day partial decay \\
E & 7-11 Mar & decay phase only (no observations during rise) \\
F & 3-20 Apr & long rise (no observations during decay) \\
G & 30May-15Jun & long rise (no observations during most of decay) \\
\tableline
\end{tabular}
\end{center}
\end{table}
\clearpage
Hardness-intensity diagrams have also been produced from the data during individual periods of flux variation in order to search for evidence of hysteresis effects, which could be used to evaluate the cooling and acceleration timescales.  Based upon the behavior of the flux, several time periods have been evaluated independently, as defined in table 1 and shown in figures 5 and 6.  The time periods evaluated in figure 5 are those that have a rise and a subsequent decay in flux observed by RXTE-PCA.  For three of these time periods, a well-defined spectral hysteresis is observed with the trace following a clockwise pattern in the representation plotted.  For the fourth (region B), there is also a consistent hardness-flux relationship observed, which tends to harden the spectrum with increasing flux, but the handedness of the HID can not be determined within the resolution of the measurements.

\clearpage
\begin{figure}
\plotone{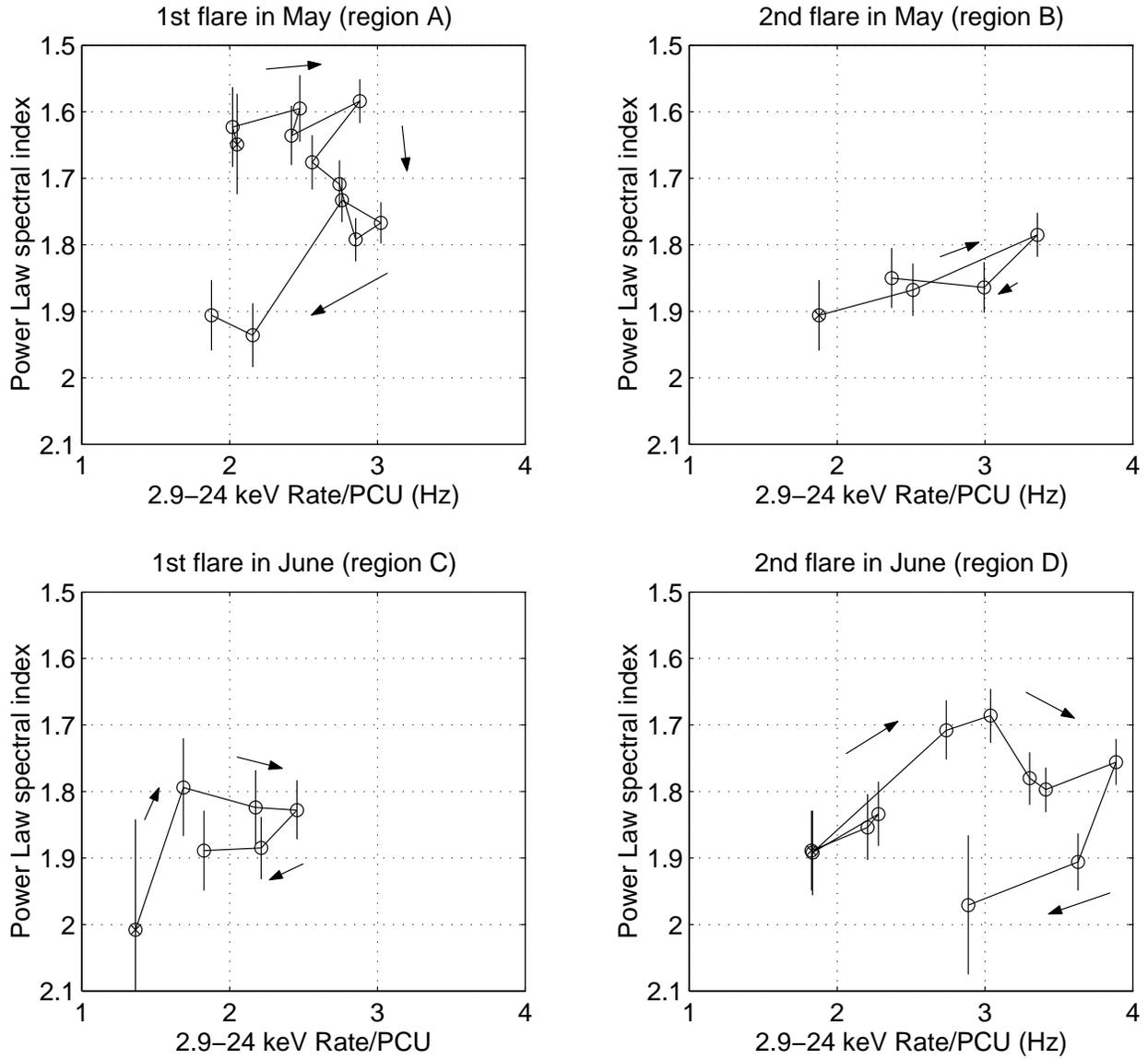}
\caption{Hardness-intensity diagrams for regions A, B, C, and D (clockwise from top-left) as defined in table 1, which all exhibit a rise and subsequent decay of flux.  Arrows have been inserted to show the direction of temporal evolution.  \label{fig5}}
\end{figure}
\clearpage
The time periods evaluated in figure 6 are those which do not have a well-defined rise and decay, but do exhibit variable flux.  Region E exhibits a fast decay, but there were no observations during the rise of the flare.  The HID exhibits a softening spectra with the decrease in flux.  Region F is characterized by a long rise in flux over a time period of $>$2 weeks, but there are no observations during the decay of flux.  This HID diagram exhibits an overall hardening of the spectrum as the flux increases, but there is some scatter in the data points.  Hysteresis properties cannot be determined due to the lack of data during the presumed decay of the flux.  Region G is similar to region F in that it is a long rise of flux over a period of $\sim$2 weeks, but it does turn over and begin to decay in the last two days.  This HID exhibits spectral hardening with increased flux.  An overall clockwise hysteresis appears to be evident for region G, however this is not well determined due to two factors; the complete long-timescale flux variation is not observed throughout the entire decay, and the short-timescale flux variations make it difficult to evaluate the characteristics of the longer timescale variability.  The two smaller HID loops associated with regions C and D flaring can also be seen in this HID.  The effect of relatively short term variability combined with long-term variability is evident in the HID diagram for region G.          
\clearpage
\begin{figure}
\plotone{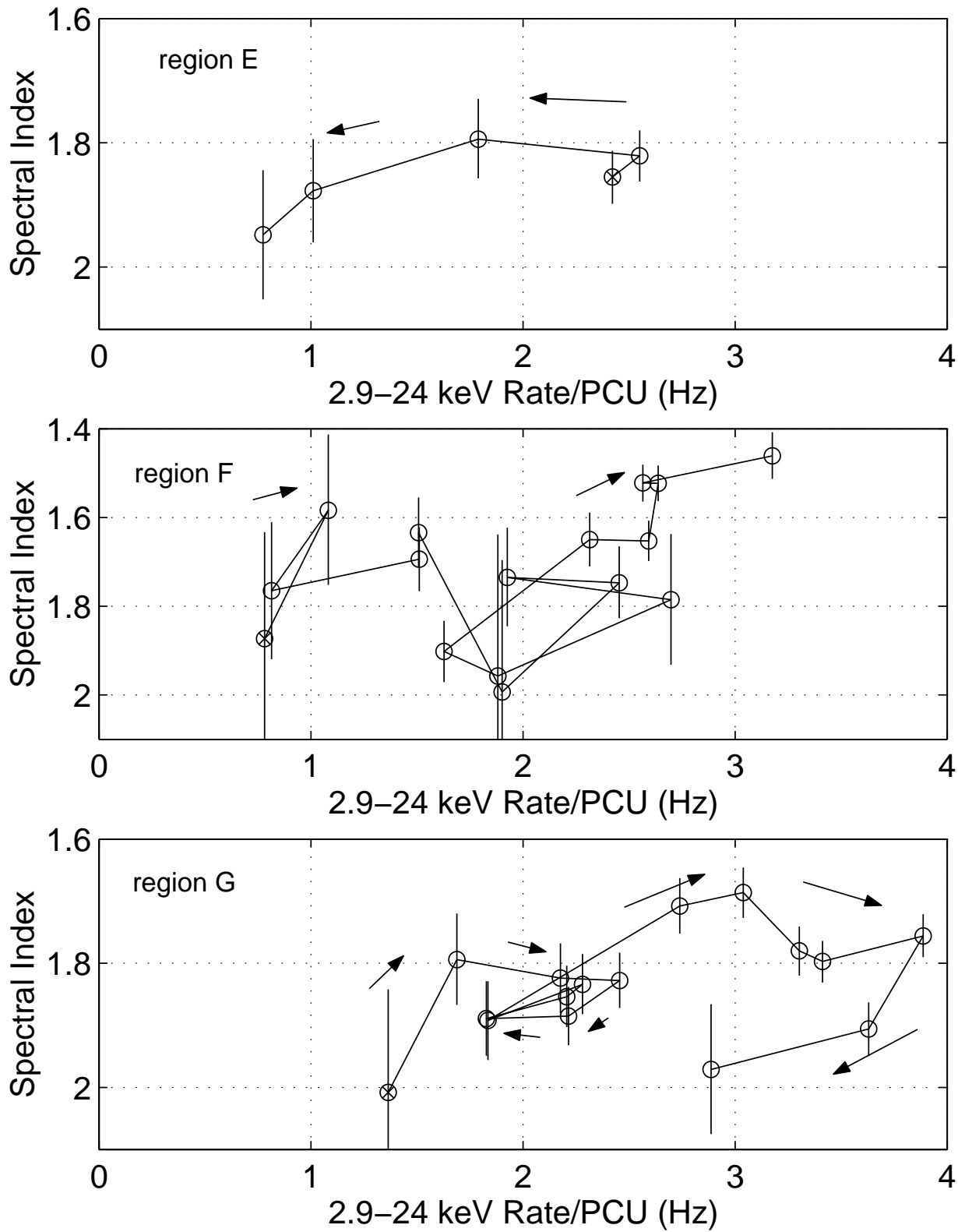}
\caption{Hardness-intensity diagrams for regions E, F, and G (left to right) as defined in table 1, which only exhibit a fraction of a flux rise/decay cycle.  Arrows have been inserted to show the direction of temporal evolution.  \label{fig6}}
\end{figure}
\clearpage
\section{Discussion and Conclusions}
This long campaign was designed to continuously monitor H1426+428 at X-ray energies (and at other wavelengths that will be reported on in a forthcoming paper) in an effort to explore variability at multiple timescales.  In spite of the fact that there were no large flares, relative to those frequently observed for other TeV emitting AGN, the campaign successfully observed variability with doubling timescales ranging from $\sim$1 day to $>$2 weeks.  In several cases, the short timescale flaring was observed on top of the long timescale gradual variations.  This type of variability complicates studies of emission mechanisms and should be considered in general, especially for studies that include non-contemporaneous multiwavelength observations.\\

Throughout the entire campaign, a simple power law with galactic absorbtion was able to fit the data well between 2.9 and 24 keV.  Prior observations performed by \citet{sam97} between 1985 and 1994 with ASCA and BBXRT led to similar spectra at some times, but at other times there were some notable differences.  At some times, the spectra from \citet{sam97} could be fit by a single power law, but at other times a broken power law that became softer at higher energies was inferred.  For the part of the spectrum between the break energy and $\sim$10 keV, the spectra had a typical power law index of $\sim$2.3 above a break energy of $\sim$2 keV for the observations described by \citet{sam97}.  This is somewhat softer than any of the spectra derived during the present campaign.  The observations reported by \citet{sam97} all show 2-10 keV fluxes that fall within the range of fluxes observed during this campaign, whereas the spectra of these historical observations are softer than the more recent observations reported in this paper and in \citet{cos01}.  This is most easily interpreted as time variability, rather than interpreting it as a contradiction between the two results, when consideration is given to the BeppoSAX observations of \citet{cos01} which also leads to a different spectral index and when consideration is given to the observed variability during this four month campaign.  It is also important to note that this has involved a comparison of different instruments with different band passes.  Another interesting feature of the BBXRT observations reported by \citet{sam97} was the evidence for a spectral line at $\approx$0.6 keV, which would imply the presence of absorbing material.  No spectral lines were observed in the 2.9 to 24 keV band covered by the observations presented in this paper (PCA can not observe down to 0.6 keV where the spectral feature was previously observed).\\

The power law spectral index was observed to vary throughout the range bounded by 1.46$\pm$0.05 and 2.03$\pm$0.09.  If one interprets this as being due to a shift in the location of the first peak of the spectral energy distribution, then it implies that the peak is sometimes in excess of 100 keV, while at other times it falls into the 2.9-24 keV region observed during this campaign.  While past observations of H1426+428 have implied a high first peak in the spectral energy distribution (SED) \citep{cos01}, there have also been observations from which a spectrum with $\Gamma$$>$2 was derived \citep{sam97}.  Based on this campaign it can be seen that significant spectral variability is present on multiple timescales.  The average spectral index was generally harder than a flat $\nu$f$_\nu$ spectrum, which implies that the first peak of the average SED was probably in excess of $\sim$100 keV (PCA alone cannot derive the actual peak location for these cases, but past observations \citep{cos01} with BeppoSAX have shown the peak to be in excess of $\sim$100 keV when the spectral index is $\sim$1.9 at PCA energies).  At other times the spectral index was flat ($\Gamma\sim$2), which implies that the first peak of the SED was in the observed 2.9-24 keV region.  It is interesting to note that this movement in the first peak of the SED during this campaign is not accompanied by any measurable fluctuations in the TeV emission from this known TeV source \citep{fal03}, but the probability of detecting such associated TeV fluctuations from such a weak source has yet to be evaluated.  The overall spectral energy distribution and its implications will be discussed in more detail in a forthcoming article that will incorporate radio, optical, X-ray, and VHE gamma-ray observations carried out contemporaneously during this campaign.\\

The longest flux variability timescale observed was the long rise and short decay in June (region G).  In spite of the fact that only a few data points were available for the decay phase, some hysteresis was evident.  The clockwise loop approached closure, as shown in the third panel of figure 6.  Although it is only marginally significant, it is interesting to note the small loop in the middle of the HID diagram, which coincides with the shorter timescale flux variability that occurred in the midst of the long rise in June.  This behavior illustrates the need for sufficient sampling over complete flaring timescales. It is probably this effect of hysteresis within hysteresis due to multiple flares, as well as flares that are non-simultaneous, that causes the scatter on the overall HID plotted in figure 4.  In spite of this scatter, it has been shown that there is a tighter relationship between flux and spectral hardness when each flux variability episode is studied independently on its proper timescale.  \\

Spectral hysteresis has been observed during several periods of flux variability during this campaign.  For an unambiguous observation of spectral hysteresis, it is necessary to observe during a large fraction of both the decay and the rise of a flare.  For three of the four cases in which this condition was met during this campaign, spectral hysteresis loops tracing a clockwise pattern were observed (given the axis orientation shown in figure 5).  In the fourth case (region B), no hysteresis and associated orientation could be observed, but the spectrum did harden as the flux increased during the rise and decay of flux.  In the past, two other high-peaked BL Lac AGN that are known TeV emitters (Mrk 421 and PKS 2155-304) have exhibited similar clockwise loops in their hysteresis plots \citep{tak96, kat00}. As mentioned previously, \citet{sem02} did not observe hysteresis with XMM observations of Mrk 421, however these observations did not sample the rise and fall of the flux variability so it is reasonable to expect a null result in that case.  The one flare that was completely observed by \citet{sem02} did display the characteristic spectral softening with decreasing flux, but no loop could be discerned from this low amplitude variation.  Another observation of Mrk 421 using BeppoSAX \citep{zha02} showed evidence of spectral hysteresis with the orientation in the counter-clockwise direction.  So, for at least one object, the hysteresis characteristics can change from flare to flare.  Since the hysteresis probes the timescale of the cooling processes as a function of energy, it appears as though some flares have a cooling time that acts faster at higher energies, consistent with synchrotron cooling, while other flares exhibit cooling times that are longer.  For the H1426+428 observations presented here, the generally clockwise hysteresis curves are consistent with the scenario in which the X-ray flux is dominated by synchrotron emission since the synchrotron cooling timescale is shorter at higher energies.  In contrast, if the hysteresis loops were oriented in the opposite direction, one would expect the source to be in a regime in which the cooling and acceleration times were approximately equal, according to models of \citet{kir98}.

By performing a study of the X-ray hysteresis, one can also explore the potential of various acceleration mechanisms to produce the TeV emission.  Typically, clockwise hysteresis patterns (using the axis orientation shown in preceding figures) are interpreted as a soft time lag in X-ray emission, while counter-clockwise orientation is interpreted as a hard lag \citep{kir98,kat00b}.  A soft lag at X-ray energies, which is interpreted as the energy dependence of the synchrotron cooling time, is characteristic of synchrotron self Compton models \citep{tak96}.  A hard lag is not consistent with some standard one-zone synchrotron self Compton models, while models such as that of \citet{kir98}, which incorporates a time-dependent propagating shock that accelerates electrons in the region of the shock front and allows for the case where $\tau_{cooling}\sim\tau_{accel}$, can be consistent with a hard lag.  The observations presented here are generally in agreement with standard models that predict a soft lag, but the lack of hysteresis from the flare in region B does not favor simple one-component models during that particular time period.  However, the model of \citet{li00} can reproduce hysteresis loops that are not clockwise (and presumably intermediate situations with no obvious loop orientation) by increasing the injection energy to the point where synchrotron self Compton losses begin to dominate the electron cooling.  The TeV emission during this entire campaign was not in a high flaring state so firm conclusions regarding TeV acceleration during any short time period cannot be made.  \\

Detailed modelling can be used to interpret the observed hysteresis patterns.  \citet{bot02} have worked on such modelling of low-peaked BL Lacs, and they have found that external Compton and synchrotron self Compton mechanisms will produce different signatures at X-ray energies. However, this study was aimed at low-peaked BL Lacs, rather than high-peaked BL Lacs.  We have found that a comparison to these models is inconclusive, probably due to the significant difference in the parameter space sampled by X-ray observations since low peaked objects are viewed in an energy region much higher with respect to the synchrotron peak, relative to high-peaked objects.  \citet{li00} did model the hysteresis characteristics of high peaked BL Lacs, but they confined themselves to only synchrotron self Compton mechanisms.  They did find that the hysteresis characteristics are sensitive to total injection energy leading them to conclude that it would be difficult to draw any conclusions based upon hysteresis observations.  However, for high-peaked BL Lac observations at energies at or below the synchrotron peak (such as the observations presented here), \cite{li00} found that clockwise orientation could be expected when the injection energy was low and synchrotron losses were dominant, and the opposite orientation could be expected when the injection energy was high and inverse Compton emission was dominant.  This supports the conclusion that for at least some of the flares observed during this campaign (those with clockwise orientation) the synchrotron losses were the dominant cooling mechanism.  More detailed modelling, which is beyond the scope of the current paper, of X-ray spectral characteristics that can be used to differentiate between various high energy emission models for high-peaked BL Lacs is needed.\\

\acknowledgments
We acknowledge the helpful service of HEASARC and the RXTE guest observer facility, particularly the helpful interaction with A. Smale and K. Jahoda.  This research is supported by grants from the U.S. Department of Energy.



\begin{thebibliography}{}

\bibitem[Aharonian (2000)]{aha00} Aharonian, F. A. 2000, New Astronomy, 5, 377

\bibitem[Blandford and Koenigl (1979)]{bla79} Blandford, R.D., and Koenigl, A. 1979, \apj, 232, 34

\bibitem[B\"{o}ttcher and Chiang (2002)]{bot02} B\"{o}ttcher, M. and Chiang, J. 2002, \apj, 581, 127

\bibitem[Buckley et al. (1996)]{buc96} Buckley, J.H., et al. 1996, \apj, 472, L9

\bibitem[Catanese and Weekes (1999)]{cat99} Catanese, M. \& Weekes, T. C. 1999, PASP, 111, 1193

\bibitem[Costamante et al. (2001)]{cos01} Costamante, L. et al. 2001, Astronomy \& Astrophysics, 371, 512

\bibitem[Dermer, Schlickeiser, \& Mastichiadis (1992)]{der92} Dermer, C.D., Schlickeiser, R., \& Mastichiadis, A. 1992, Astron. \& Astrophys., 256, L27

\bibitem[Falcone et al. (2003)]{fal03} Falcone, A., et al. 2003, to be published in New Astronomy Reviews

\bibitem[Fossati et al. (1998)]{fos98} Fossati, G. et al. 1998, Monthly Notices of the Royal Astronomical Society, 299, 433 

\bibitem[Ghisellini et al. (1998)]{ghi98} Ghisellini, G. et al. 1998, Monthly Notices of the Royal Astronomical Society, 301, 451 

\bibitem[Hartman et al. (1999)]{har99} Hartman, R. C., et al. 1999, \apj, 123, 79

\bibitem[Holder et al. (2003)]{hol03} Holder, J., at al. 2003, \apjl, 583, L9

\bibitem[Horan et al. (2002)]{hor02} Horan, D., et al. 2002, \apj, 571, 753

\bibitem[Kataoka et al. (2000)]{kat00} Kataoka, J., Takahashi, T., Makino, F., Inoue, S., Madejski, G.M., Tashiro, M., Urry, C.M., and Kubo, H. 2000, \apj, 528, 243

\bibitem[Kataoka (2000)]{kat00b} Kataoka, J., Ph.D. Thesis, Univ. Tokyo  

\bibitem[Kirk, Rieger, and Mastichiadis (1998)]{kir98} Kirk, J.G., Rieger, F.M., and Mastichiadis, A. 1998, Astron. \& Astrophys., 333, 452

\bibitem[Kniffen et al. (1993)]{kni93} Kniffen, D.A., et al. 1993, \apj, 411, 133

\bibitem[Koenigl (1981)]{koe81} Koenigl, A. 1981, \apj, 243, 700

\bibitem[Li and Kusunose (2000)]{li00} Li, H. and Kusunose, M. 2000, \apj, 536, 729

\bibitem[Macomb et al. (1995)]{mac95} Macomb, D.J., et al. 1995, \apjl, 449, L99

\bibitem[Madejski et al. (1992)]{mad92} Madejski, G.M., et al. 1992, in Proc. of 28th Yamada Conf.: Frontiers of X-ray Astronomy, ed. Y. Tanaka \& K. Koyama (Tokyo: Universal Academy Press), 583

\bibitem[Mannheim (1993)]{man93} Mannheim, K. 1993, Astron. \& Astrophys., 269, 67 

\bibitem[Mattox et al. (1993)]{mat93} Mattox, J.R., et al. 1993, \apj, 410, 609 

\bibitem[Morrison and McCammon (1983)]{mor83} Morrison,R., and McCammon, D. 1983, \apj, 270, 119

\bibitem[Muecke et al. (2003)]{mue03} Muecke, A., Protheroe, R.J., Engel, R., Rachen, J.P., \& Stanev, T. 2003, Astroparticle Physics, 18, 593

\bibitem[Muecke and Protheroe (2001)]{mue01} Muecke, A. \& Protheroe, R. J. 2001, Astroparticle Physics, 15, 121

\bibitem[Petry et al. (2002)]{pet02} Petry, D. et al. 2002, \apj, 580, 104

\bibitem[Pian et al. (1998)]{pia98} Pian, E. et al. 1998, \apjl, 492, L17  

\bibitem[Quinn et al. (1996)]{qui96} Quinn, J., et al. 1996, \apj, 456, L83 

\bibitem[Remillard et al. (1989)]{rem89} Remillard, R., Tuohy, I.R., Brissenden, R.J.V., Buckley, D.A., Schwartz, D.A., Feigelson, E.D., and Tapia, S. 1989, \apj, 345, 140

\bibitem[Sambruna et al. (1997)]{sam97} Sambruna, R., George, I. M., Madejski, G., Urry, C.M., Turner, T. J., Weaver, K.A., Maraschi, L., Treves, A. 1997, \apj, 483, 774

\bibitem[Sembay et al. (2002)]{sem02} Sembay, S., Edelson, R., Markowitz, A., Griffiths, R.G., and Turner, M.J.L. 2002, \apj, 574, 634

\bibitem[Sikora, Begelman, \& Rees (1994)]{sik94} Sikora, M., Begelman, M.C., \& Rees, M.J. 1994, \apj, 421, 153  
\bibitem[Stark et al. (1992)]{sta92} Stark, A.A., Gammie, C.F., Wilson, R.W., Bally, J., Linke, R.A., Heiles, C. and Hurwitz, M. 1992, \apjs, 79, 77

\bibitem[Takahashi et al. (1996)]{tak96} Takahashi, T., et al. 1996, \apjl, 470, L89

\bibitem[von Montigny et al. (1995)]{von95} von Montigny, C., et al. 1995, \apj, 440, 525 

\bibitem[Zhang (2002)]{zha02} Zhang, Y.H. 2002, Monthly Notices of the Royal Astronomical Society, 337 609

\end{thebibliography}
\end{document}